\documentclass{article} 
\usepackage{graphicx}
\usepackage{hyperref}
\usepackage{indentfirst}
\date{\vspace{-5ex}} 
\title{Millisecond pulsars as standards:\\ Timing, positioning and communication}
\author{Cl\'ement Vidal}

\begin{document}
\renewcommand{\abstractname}{\vspace{-\baselineskip}} 

\noindent
\begin{footnotesize}
\\ To appear in \textit{Proceedings IAU Symposium} 337, Pulsar Astrophysics -­ The Next 50 Years, edited by P. Weltevrede, B. B. P. Perera, L. Levin Preston, and S. Sanidas. Jodrell Bank Observatory, UK, 2017.
\end{footnotesize}

\begingroup
\let\newpage\relax
\maketitle
\endgroup
\begin{center} Center Leo Apostel, Vrije Universiteit Brussel, \\ Krijgskundestraat 33, 1160 Brussels, Belgium \\contact@clemvidal.com\\
\end{center} 
\begin{abstract}
\centering\begin{minipage}{\dimexpr\paperwidth-12cm}

{\textit{Abstract}}: Millisecond pulsars (MSPs) have a great potential to set standards in timekeeping, positioning and metadata communication. \\

\end{minipage}
\end{abstract}

\section{Timekeeping: why use millisecond pulsars as clocks?}

A pulsar timescale provides a parallel standard to terrestrial ones, that is based on macroscopic neutron stars behaviour instead of quantum processes. Because it is not based on Earth, and distributed amongst millisecond pulsars in the galaxy, it is robust to any catastrophic risk on Earth. MSPs are long-lived, so a pulsar timescale will remain operational longer than any clock we can construct on Earth (see e.g. Hobbs et al 2012).

The usefulness of a pulsar timescale is likely to rise, as international collaborations between various pulsar timing arrays will make their precision improve dramatically (Manchester et al 2017).

\section{Positioning: why navigate with millisecond pulsars?}

MSPs have also been shown to be useful for galactic navigation, as they provide all the necessary ingredients for a passive and accurate galactic positioning system. This is known in astronautics as X-ray pulsar-based navigation (XNAV, see e.g. Sheikh et al. 2006; Becker, Bernhardt, and Jessner 2013). XNAV uses a time-of-arrival navigation method comparable to GPS, accurate down to about 100 meters. MSPs thus promise to constitute a positioning and navigation standard for future navigation in the solar system and in the galaxy.

As with EarthÕs clocks and GPS, timing and positioning standards have many more applications beyond clocks and navigation, such as the general synchronization of actions. With MSPs, a similarly wide array of applications might hold at a galactic scale. 

\section{Communication: why use millisecond pulsars for metadata?}

Humans have already used pulsars to locate the Earth, back in 1972 with the plaque attached to the \textit{Pioneer 10} spacecraft (Sagan, Sagan and Drake 1972).

If we assume that there are other civilisations in the galaxy, MSPs could be key in providing metadata standards for any communication. Any letter or email contains metadata information about where it comes from, where it goes, and when it was written. We can expect that similar conventions exist for any potential galactic communication. Interstellar messages are likely to be galacto-tagged and pulsar-time-stamped by reference to MSPs. This is consistent with the notion of \textit{astrophysical coding} that suggests to encode interstellar messages according to a shared astrophysical context, such as pulsars (Cordes and Sullivan 1995, Sullivan and Cordes 1995).

This simple remark is constraining for SETI. Given any suspicious message we find, the first step becomes to attempt to decode not the message itself, but its metadata (Vidal 2017, sec. 5). It's much easier to decode metadata information, than encoded information.

\section{Little Green Men after all?}

Could extraterrestrial civilisations modulate pulsar signals (Chennamangalam et al. 2015) to maintain a Pulsar Positioning System (Vidal 2017)? I suggested lines of inquiry to test this idea, that may lead to new predictions regarding the spatial and power distribution of MSPs in the galaxy; their population; their evolutionary tracks; possible synchronisation between MSPs; as well as decoding metadata or information in MSPs' pulses. Such an instance of stellar engineering operating on binary stars (Vidal 2016) would be an example of a Type II civilisation on Kardashev's (1964) scale. 
\newpage
\section{Conclusion}

Millisecond pulsars hold great promises for the future of humanity, to set timekeeping, positioning and metadata communication standards.

Astrophysicists already use the exceptional timing properties of MSPs for many purposes and more may be discovered. Astronautics engineers are testing pulsar navigation now, with the XPNAV1 satellite (China) and NASA's NICER mission (USA). For potential communication with ETI, we have already used pulsar metadata to message to potential extraterrestrial intelligence, and such metadata is arguably a simple kind of signal we might be able to decipher.

\begin{footnotesize}
\section*{References}
\hspace{0pt} Becker, W., M. G. Bernhardt, \& A. Jessner. 2013. Autonomous Spacecraft Navigation With Pulsars.
\textit{Acta Futura} 7: 11-28. doi:10.2420/AF07.2013.11. \\  \url{http://arxiv.org/abs/1305.4842}.

Chennamangalam, J., A. Siemion, D. R. Lorimer, \& D. Werthimer. 2015. Jumping the Energetics Queue. \textit{New Astron.} 34: 245-249. \url{http://arxiv.org/abs/1311.4608}.

Cordes, J. M., and W. T. III Sullivan. 1995. Astrophysical Coding: A New Approach to SETI Signals. I. In \textit{Progress in the Search for Extraterrestrial Life}, 74:325-34. ASP Conference Series. 

Hobbs, G., et al. 2012. Development of a Pulsar-Based Time-Scale.
 \textit{MNRAS} 427 (4): 2780-87.

Kardashev, N. S. 1964. Transmission of Information by Extraterrestrial Civilizations. 
\textit{Soviet Astron.} 8 (2): 217-220. \url{http://adsabs.harvard.edu/abs/1964SvA.....8..217K}.

Manchester, R. N., et al. 2017. Pulsars: Celestial Clocks. In 
\textit{The Science of Time 2016}, 253-65. Astrophysics and Space Science Proceedings. Springer. doi:10.1007/978-3-319-59909-0\_30.

Sagan, C., L. S. Sagan, \& F. Drake. 1972. A Message from Earth.
 \textit{Science} 175 (4024): 881-84.

Sheikh, S. I., D. J. Pines, P. S. Ray, K. S. Wood, M. N. Lovellette, \& M. T. Wolff. 2006.
Spacecraft Navigation Using X-Ray Pulsars. \textit{Journal of Guidance, Control, and Dynamics} 29 (1): 49-63. doi:10.2514/1.13331.

Sullivan, W. T. III, \& J. M. Cordes. 1995. Astrophysical Coding: A New Approach to SETI Signals. II. In \textit{Progress in the Search for Extraterrestrial Life}, 74:337-42. ASP Conference Series.

Vidal, C. 2016. Stellivore Extraterrestrials? Binary Stars as Living Systems. 
\textit{AcA} 128:251-56. doi:10.1016/j.actaastro.2016.06.038. \url{http://zenodo.org/record/164853}.

Vidal, C. 2017. Pulsar Positioning System: A Quest for Evidence of Extraterrestrial Engineering.
\textit{International Journal of Astrobiology}, doi:10.1017/S147355041700043X. \\ \url{http://arxiv.org/abs/1704.03316}.

\end{footnotesize}
\end{document}